\documentclass[aps,preprint,showpacs,superscriptaddress]{revtex4}
\newcommand{\beq}{\begin{equation}}
\newcommand{\eeq}{\end{equation}}
\begin{document}

\title{Dissymmetrical
tunnelling in heavy fermion metals}

\author{V.R.  Shaginyan \footnote{E--mail:
vrshag@thd.pnpi.spb.ru}} \affiliation{Petersburg Nuclear Physics
Institute, Russian Academy of Sciences,  Gatchina, 188300, Russia}
\affiliation{ CTSPS, Clark Atlanta University, Atlanta, Georgia
30314, USA}

\begin{abstract}

A tunnelling conductivity between a heavy fermion metal and a
simple metallic point is considered. We show that at low
temperatures this conductivity can be noticeably dissymmetrical
with respect to the change of voltage bias. The dissymmetry can be
observed in experiments on the heavy fermion metals whose
electronic system has undergone the fermion condensation quantum
phase transition.

\end{abstract}


\maketitle

Understanding the unusual quantum critical properties of
heavy-fermion (HF) metals at low temperatures $T$ remains
challenging. It is a common belief that  quantum phase transitions
developing in the HF metals at $T=0$, which have ability to
influence the finite temperature properties,  are responsible for
the anomalous behavior. Experiments on the HF metals explore
mainly their thermodynamic properties which proved to be quite
different from that of ordinary metals described by the Landau
Fermi liquid (LFL) theory. In the LFL theory, considered as the
main instrument when investigating quantum many electron physics,
the effective mass $M^*$ of quasiparticle excitations controlling
the density of states determines the thermodynamic properties of
electronic systems. It is possible to explain the observed
thermodynamic properties of the HF metals on the basis of the
fermion condensation quantum phase transition (FCQPT) which allows
the existence of the Landau quasiparticles down to the lowest
temperatures \cite{shag4,ckhz}. In contrast to the Landau
quasiparticles, these are characterized by the effective mass
which  strongly depends on temperature $T$, applied magnetic field
$B$ and the number density $x$ of the heavy electron liquid of HF
metal. Thus, we come back again to the key role of the density of
state. It would be desirable to probe the other properties of the
heavy electron liquid such as the probabilities of quasiparticle
occupations which are not directly linked to the density of states
or to the behavior of $M^*$. Scanning tunnelling microscopy (STM)
being sensitive to both the density of states and the
probabilities of quasiparticle occupations is an ideal technique
for the study of such effects at quantum level.

The tunnelling current $I$ through the point contact between two
ordinary metals is proportional to the driving voltage $V$ and to
the squared modulus of the quantum mechanical transition amplitude
$t$ multiplied by the difference
$N_1(0)N_2(0)(n_1(p,T)-n_2(p,T))$, see e.g. \cite{zag}. Here
$n(p,T)$ is the quasiparticle distribution function and $N(0)$ is
the density of states of the corresponding metal. On the other
hand, the wave function calculated in the WKB approximation and
defining $t$ is proportional to $(N_1(0)N_2(0))^{-1/2}$. As a
result, the density of states is dropped out  and the tunnelling
current does not depend on $N_1(0)N_2(0)$. Upon taking into
account that at $T\to 0$ the distribution $n(p,T\to0)\to n_F(p)$,
where $n_F(p)$ is the step function $\theta(p-p_F)$ with $p_F$
being the Fermi momentum, one can check that within the LFL theory
the differential tunnelling conductivity $\sigma_d(V)=dI/dV$ is a
symmetric function of the voltage $V$. In fact, the symmetry of
$\sigma_d(V)$ holds provided that so called particle-hole symmetry
is preserved as it is within the LFL theory. Therefore, the
existence of the $\sigma_d(V)$ symmetry is quite obvious and
common in the case of metal-to-metal contacts when these metals
are in the normal state or in the superconducting one.

In this letter we show that the situation can be different when
one of the two metals is a HF metal whose electronic system is
represented by the heavy electron liquid. When the heavy electron
liquid has undergone FCQPT its distribution function is no longer
the step function as soon as the temperature tends to zero
\cite{ks}. As a result, both the differential tunnelling
conductivity $\sigma_d(V)$ and the tunnelling conductivity
$\sigma(V)$ become dissymmetrical as a function of voltage $V$.
While the application of magnetic field destroying the non-Fermi
liquid (NFL) behavior of the heavy electron liquid restores the
symmetry.

At first, we briefly describe the heavy electron liquid with the
fermion condensate (FC) \cite{ks,ksk,shag1}. When the number
density $x$ of the liquid approaches some density $x_{FC}$ the
effective mass diverges. Because the kinetic energy near the Fermi
surface is proportional to the inverse effective mass,  FCQPT is
triggered by the frustrated kinetic energy. Behind the critical
point $x_{FC}$, the quasiparticle distribution function
represented by $n_F(p)$ does not deliver the minimum to the Landau
functional $E[n({\bf p})]$. As a result, at $x<x_{FC}$ the
quasiparticle distribution is determined by the standard equation
to search the minimum of a functional \cite{ks}
\begin{equation} \frac{\delta E[n({\bf p})]}{\delta n({\bf
p},T=0)}=\varepsilon({\bf p})=\mu; \,p_i\leq p\leq p_f.
\end{equation}
Equation (1) determines the quasiparticle distribution function
$n_0({\bf p})$ which delivers the minimum  value to the ground
state energy $E$. Being determined by Eq. (1), the function
$n_0({\bf p})$ does not coincide with the step function $n_F(p)$
in the region $(p_f-p_i)$, so that $0<n_0({\bf p})<1$, while
outside the region it coincides with $n_F(p)$. It follows from Eq.
(1) that the single particle spectrum is completely flat over the
region. Such a state was called the state with FC because
quasiparticles located in the region $(p_f-p_i)$ of momentum space
are pinned to the chemical potential $\mu$.  We note that the
behavior obtained as observed within exactly solvable models
\cite{irk,lid} and represents a new state of Fermi liquid
\cite{vol}. We can conclude that the relevant order parameter
$\kappa({\bf p})=\sqrt{n_0({\bf p})(1-n_0({\bf p}))}$ is the order
parameter of the superconducting state with the infinitely small
value of the superconducting gap $\Delta$ \cite{ksk}. Thus this
state cannot exist at any finite temperatures and driven by the
parameter $x$: at $x>x_{FC}$ the system is on the disordered side
of FCQPT; at $x=x_{FC}$, Eq. (1) possesses the non-trivial
solutions $n_0({\bf p})$ with $p_i=p_F=p_f$; at $x<x_{FC}$, the
system is on the ordered side. At $T>0$, the quasiparticle
distribution is given by
\begin{equation} n({\bf p},T)=\left\{ 1+\exp
\left[ \frac{(\varepsilon({\bf p},T)-\mu)}T \right] \right\}
^{-1},
\end{equation}
where $\varepsilon({\bf p},T)$ is the single-particle spectrum, or
dispersion, of the quasiparticle excitations and $\mu$ is the
chemical potential. Equation (2) can be recast as
\begin{equation} \varepsilon ({\bf p},T)-\mu (T)=T\ln \frac{1-n({\bf
p},T)}{n({\bf p},T)}.  \end{equation} As $T\to 0$, the logarithm
on the right hand side of Eq. (3) is finite when $p$ belongs to
the region $(p_f-p_i)$, therefore $T\ln(...)\to 0$, and we again
arrive at Eq. (1). Near the Fermi level the single particle
spectrum can be approximated as \beq\varepsilon(p\simeq
p_F,T)-\mu\simeq \frac{p_F(p-p_F)}{M^*}.\eeq It follows from Eq.
(2) that $n(p,T\to 0)\to n_F(p)$ provided that $M^*$ is finite at
$T\to 0$. Thus at low temperatures, the left hand side of Eq. (3)
determines the behavior of the right hand side. In contrast to
this case, the right hand side of Eq. (3) determines the behavior
of $M^*$ when FC is set in at the liquid. Indeed, it follows from
Eq. (1) that $n({\bf p},T\to0)=n_0({\bf p})$. Therefore at low
temperatures, as seen from Eq. (3), the effective mass diverges as
\cite{ams} \beq M^*(T)\simeq p_F\frac{p_f-p_i}{4T}.\eeq At $T\ll
T_f$, Eq. (5) is valid and determines quasiparticles with the
energy $z$ and characterized by the distribution function
$n_0(p)$. Here $T_f$ is the temperature at which the influence of
FCQPT vanishes \cite{ksk}. The energy $z$  belongs to the interval
\beq\mu-2T\leq z \leq\mu+2T.\eeq

Now we turn to a consideration of the tunnelling current at low
temperatures which in the case of ordinary metals is given by
\cite{zag} \beq
I(V)=2|t|^2\int\left[n_F(z-\mu)-n_F(z-\mu+V)\right]dz.\eeq We use
an atomic system of units: $e=m=\hbar =1$, where $e$ and $m$ are
electron charge and mass, respectively. Since temperatures are low
we approximate the distribution function of ordinary metal by the
step function $n_F$. It follows from Eq. (7) that quasiparticles
with the energy $z$, $\mu-V\leq z\leq \mu$, contribute to the
current, while $\sigma_d(V)\simeq 2|t|^2$ is a symmetrical
function of $V$. In the case of the heavy electron liquid with FC,
the tunnelling current are found to be of the form \beq
I(V)=2\int\left[n_0(z-\mu)-n_F(z-\mu+V)\right]dz.\eeq Here we have
replaced the distribution function of ordinary metal by $n_0$
being the solution of Eq. (1). We have also taken units such that
$|t|^2=1$. Assume that $V$ satisfies the condition, $|V|\leq 2T$,
while the current flows from the HF metal to ordinary one.
Quasiparticles of the energy z, $\mu-V\leq z$, contribute to
$I(V)$, and the differential conductivity $\sigma_d(V)\simeq
2n_0(z\simeq \mu-V)$. If the sign of the voltage is changed, the
direction of the current is also changed. In that case,
quasiparticles of the energy z, $\mu+V\geq z$, contribute to
$I(V)$, and the differential conductivity $\sigma_d(-V)\simeq
2(1-n_0(z\simeq \mu+V))$. The dissymmetrical part $\Delta
\sigma_d(V)=(\sigma_d(-V)-\sigma_d(V))$ of the differential
conductivity is of the form \beq \Delta \sigma_d(V)\simeq
2[1-(n_0(z-\mu\simeq V)+n_0(z-\mu\simeq -V))].\eeq It is worth
noting that it follows from Eq. (9) that  $\Delta \sigma_d(V)=0$
if the HF metal in question is replaced by an ordinary metal.
Indeed, the effective mass is finite at $T\to0$, then $n_0(T\to
0)\to n_F$ being given by Eq. (2), and $1-n(z-\mu\simeq
V)=n(z-\mu\simeq -V)$. One might say that the dissymmetrical part
vanishes due to the particle-hole symmetry. On the other hand,
there are no reasons to expect that $(1-n_0(z-\mu\simeq
V)-n_0(z-\mu\simeq -V))=0$. Thus, we are led to the conclusion
that the differential conductivity becomes a dissymmetrical
function of the voltage. To estimate $\Delta \sigma_d(V)$, we
observe that this is zero when $V=0$, because $n_0(p=p_F)=1/2$ as
it should be and it follows from Eq. (3) as well. It is seen from
Eq. (9) that $\Delta \sigma_d(V)$ is an even function of $V$.
Therefore we can assume that at low values of the voltage $V$ the
dissymmetrical part behaves as $\Delta \sigma_d(V)\propto V^2$.
Then, the natural scale to measure the voltage is $2T$ as it is
seen from Eq. (6). In fact, the dissymmetrical part is to be
proportional to $(p_f-p_i)/p_F$. As a result,  we obtain \beq
\Delta \sigma_d(V)\simeq
c\left(\frac{V}{2T}\right)^2\frac{p_f-p_i}{p_F}.\eeq Here $c$ is a
constant which is expected to be of the order of unit. This
constant can be evaluated by using analytical solvable models. For
example, calculations of $c$ within a simple model, when the
Landau functional $E[n(p)]$ is of the form \cite{ks} \beq
E[n(p)]=\int \frac{p^2}{2M}\frac{d{\bf p}}{(2\pi)^3}+V_1\int
n(p)n(p)\frac{d{\bf p}}{(2\pi)^3},\eeq give that $c\simeq 1/2$. It
follows from Eq. (10) that when $V\simeq 2T$ and FC occupies a
noticeable part of the Fermi volume, $(p_f-p_i)/p_F\simeq 1$, the
dissymmetrical part becomes comparable with differential
tunnelling conductivity, $\Delta \sigma_d(V)\sim  V_d(V)$.

The dissymmetrical behavior of the tunnelling conductivity can be
observed in measurements on the heavy fermion metals, for example,
such as YbRh$_2$(Si$_{0.95}$Ge$_{0.05}$)$_2$ or YbRh$_2$Si$_2$
which are expected to have undergone FCQPT. In that case, upon the
application of magnetic field $B$ the effective mass is to diverge
as  \cite{shag4,shag} \beq M^*(B)\propto (B-B_{c0})^{\alpha}.\eeq
Here $B_{c0}$ is the critical magnetic field which drives the HF
metal to its magnetic field tuned quantum critical point. The
value of the critical exponent $\alpha=-1/2$ is in good agreement
with experimental observations collected on these metals
\cite{geg,cust}. The measurements of $\Delta \sigma_d(V)$ have to
be carried out applying magnetic field $B_{c0}$ at temperatures
$T\leq T_f$. In the case of these metals, $T_f$ is of the order of
few Kelvin \cite{shag}. We note that at sufficiently low
temperatures, the application of magnetic field $B>B_{c0}$ leads
to the restoration of the Landau Fermi liquid with $M^*(B)$ given
by Eq. (12) \cite{shag4,shag}. As a result, the dissymmetrical
behavior of the tunnelling conductivity vanishes.

The dissymmetrical differential conductivity $\Delta \sigma_d(V)$
can also be observed when the HF metal in question goes from
normal to superconducting. The reason is that $n_0(p)$ is again
responsible for the dissymmetrical part of $\sigma_d(V)$. This
$n_0(p)$ is not appreciably disturbed by the pairing interaction
which is relatively weak as compared to the Landau interaction
forming the distribution function $n_0(p)$ \cite{ams,amars}. In
the case of superconductivity, we have to take into account that
the density of states, \beq \frac{N_s(E)}{N(0)}=
\frac{|E|}{\sqrt{E^2-\Delta^2}},\eeq comes into the play because
$N_s$ is zero in the gap, that is when $|E|\leq |\Delta|$. Here
$E$ is the quasiparticle energy, while the normal state
quasiparticle energy is $\varepsilon-\mu=\sqrt{E^2-\Delta^2}$. Now
we can arrange Eq. (9) for the case of superconducting HF metal by
multiplying the right hand side of Eq. (9) by $N_s/N(0)$ and
replacing the quasiparticle energy $z-\mu$ by
$\sqrt{E^2-\Delta^2}$ with $E$ being represented by the voltage
$V$. As a result, Eq. (10) can be cast into the following form
\beq \Delta \sigma_d(V)\simeq
\frac{\left(\sqrt{V^2-\Delta^2}\right)^2}
{|\Delta|\sqrt{V^2-\Delta^2}}
\frac{p_f-p_i}{p_F}=\sqrt{\left[\frac{V} {\Delta}\right]^2-1}\,\,
\frac{p_f-p_i}{p_F}.\eeq Note that the scale $2T$ entering Eq.
(10) is replaced by the scale $\Delta$ in Eq. (14). In the same
way, as Eq. (10) is valid up to $V\simeq 2T$, Eq. (14) is valid up
to $V\simeq 2|\Delta|$. It is seen from Eq. (14) that the
dissymmetrical part of the differential tunnelling conductivity
becomes  as large as the differential tunnelling conductivity at
$V\simeq 2|\Delta|$ provided that FC occupies a large part of the
Fermi volume, $(p_f-p_i)/p_F\simeq 1$. In the case of a $d$-wave
gap, the right hand side of Eq. (14) has to be integrated over the
gap distribution. As a result, $\Delta \sigma_d(V)$ is expected to
be finite even at $V=\Delta_1$, where $\Delta_1$ is the maximum
value of the $d$-wave gap. A detailed consideration of the
superconducting case will be published elsewhere.

In summary, we have shown that the differential tunnelling
conductivity between metallic point and an ordinary metal which is
commonly symmetric as a function of the voltage becomes noticeably
dissymmetrical when the ordinary metal is replaced by a HF metal
the electronic system of which has undergone FCQPT. This
dissymmetry can be observed when the HF metal is both normal and
superconducting. We have also discussed possible experiments to
study the dissymmetry.\\

I am grateful to Dr. A.Z. Msezane for the kind hospitality at
CTSPS, Clark Atlanta University, Atlanta, GA, USA, where a  part
of this work was done. This work was supported in part by the
Russian Foundation for Basic Research.


\begin{thebibliography}{99}

\bibitem{shag4}  V.R. Shaginyan, JETP Lett. {\bf 79}, 286 (2004).

\bibitem{ckhz} J.W. Clark, V.A. Khodel, and M.V. Zverev,
Phys. Rev. B {\bf 71}, 012401 (2005).

\bibitem{zag} A.M. Zagoskin, {\it Quantum Theory of Many-Body Systems}
(Springer-Verlag New York, Inc., 1998).

\bibitem{ks}  V.A. Khodel and V.R. Shaginyan,
JETP Lett. {\bf 51}, 553 (1990).

\bibitem{ksk} V.A. Khodel, V.R. Shaginyan, and
V.V. Khodel, Phys. Rep. {\bf 249}, 1 (1994); V.A. Khodel and V.R.
Shaginyan, Condens. Matter Theories {\bf 12}, 222 (1997); J.
Dukelsky et al., Z. Phys. B {\bf 102}, 245 (1997).

\bibitem{shag1}  V.R. Shaginyan, JETP Lett. {\bf 77}, 99 (2003); V.R.
Shaginyan, JETP Lett. {\bf 77}, 178 (2003).

\bibitem{irk}  V.Yu. Irkhin, A.A. Katanin, and
M.I. Katsnelson, Phys. Rev. Lett. {\bf 89}, 076401 (2002).

\bibitem{lid}  D. Lidsky, J. Shiraishi, Y. Hatsugai, and M. Kohmoto,
Phys. Rev. B {\bf 57}, 1340 (1998).

\bibitem{vol} G. E. Volovik, JETP Lett. {\bf 53}, 222 (1991).

\bibitem{ams} M.Ya. Amusia and V.R. Shaginyan, Phys. Rev. B {\bf 63},
224507 (2001).

\bibitem{shag}  Yu.G. Pogorelov and V.R. Shaginyan,
JETP Lett. {\bf 76}, 532 (2002); V.R. Shaginyan, A.Z. Msezane, and
M.Ya. Amusia, cond-mat/0501093.

\bibitem{geg} P. Gegenwart {\it et al}., Phys. Rev. Lett.
{\bf 89}, 056402 (2002).

\bibitem{cust} J. Custers {\it et al}., Nature {\bf 424}, 524 (2003).

\bibitem{amars} M.Ya. Amusia, S.A. Artamonov, and V.R. Shaginyan,
JETP Lett. {\bf 74}, 396 (2001).

\end{thebibliography}
\end{document}